# Thermodynamics of Radiation Pressure and Photon Momentum (Part 2)

Masud Mansuripur

College of Optical Sciences, The University of Arizona, Tucson, Arizona, USA

**Abstract**. We derive some of the properties of blackbody radiation using thermodynamic identities. A few of the results reported earlier (in Part 1 of the present paper) will be re-derived here from a different perspective. We argue that the fluctuations of thermal radiation can be expressed as the sum of two contributions: one resulting from classical considerations in the limit when Planck's constant $\hbar$ goes to zero, and a second one that is rooted in the discrete nature of electromagnetic particles (photons). Johnson noise and the Nyquist theorem will be another topic of discussion, where the role played by blackbody radiation in generating noise within an electronic circuit will be emphasized. In the context of thermal fluctuations, we also analyze the various sources of noise in photodetection, relating the statistics of photo-electron counts to energy-density fluctuations associated with blackbody radiation. The remaining sections of the paper are devoted to an analysis of the thermodynamic properties of a monatomic gas under conditions of thermal equilibrium. This latter part of the paper aims to provide a basis for comparisons between a photon gas and a rarefied gas of identical rigid particles of matter.

**1. Introduction**. In a previous paper [1] we discussed the properties of blackbody radiation in general terms, emphasizing those aspects of the theory that pertain specifically to radiation pressure and photon momentum. In the present paper, we elaborate some of the ideas discussed in [1], derive some of the previously reported results from a different perspective, and compare the properties of a photon gas in thermal equilibrium (i.e., with the walls of an otherwise empty cavity) with the thermodynamic properties of a monatomic gas in thermal equilibrium with a reservoir at a given temperature $T$. In Sec.2 we derive the relation between the Helmholtz free energy and the partition function for a closed system containing a fixed number of particles. Section 3 is devoted to a derivation of some of the fundamental properties of thermodynamic systems under thermal equilibrium conditions. The thermodynamic identities obtained in these early parts of the paper will be needed in subsequent sections. Section 4 provides a brief derivation of the frequency at which the blackbody radiation density reaches its peak value. In Sec.5 we analyze the fluctuations of blackbody radiation in the classical limit when Planck's constant $\hbar$ goes to zero; the remaining term in the expression of the variance of blackbody radiation — that due to the discrete nature of photons — is examined in Sec.6. Sections 7 and 8 are devoted to a discussion of Johnson noise and the Nyquist theorem, where the prominent role of blackbody radiation in generating thermal noise within electronic circuits is emphasized. In Sec.9 we return to the topic of photon-number variance and examine the fluctuations of blackbody radiation in the context of idealized photodetection, relating the fluctuations of thermal light to the statistics of photodetection in the presence of chaotic illumination. The remaining sections of the paper are devoted to the thermodynamics of a simple monatomic gas under conditions of thermal equilibrium. We derive the Sackur-Tetrode expression for the entropy of a monatomic gas, and proceed to discuss the various thermodynamic properties that can be extracted from the Sackur-Tetrode equation. The goal of the latter part of the paper is to provide a basis for comparisons between the thermodynamics of the photon gas and those of a rarefied gas consisting of rigid, identical, material particles.

**2. The concept of free energy in thermodynamics**. The Helmholtz free energy $F$ of an $N$-particle system having, in equilibrium, internal energy $U$, entropy $S$, absolute temperature $T$, pressure $p$, and volume $V$ is defined as $F = U - TS$ [2−4]. An important application of the concept of free energy is found in reversible isothermal expansion (or contraction) processes. Here the heat $\Delta Q$ is given to the system while its volume expands by $\Delta V$, delivering an amount of work $\Delta W = p\Delta V$ to the outside world. The change of entropy in this process is $\Delta S = \Delta Q/T$, while the internal energy of the system changes by $\Delta U = \Delta Q - p\Delta V$. The pressure may thus be written as $p = -(\Delta U - T\Delta S)/\Delta V$, or, considering that the temperature is kept constant, as $p = -(\Delta F/\Delta V)_{T,N}$. The pressure in a reversible isothermal expansion (or contraction) is, therefore, given by $p = -(\partial F/\partial V)_{T,N}$. This expression should be contrasted with the formula $p = -(\partial U/\partial V)_{S,N}$, which is the pressure of a system undergoing a reversible adiabatic expansion (or contraction).

An intimate connection exists between the partition function $Z(T,N,V)$ of an $N$-particle thermodynamic system and its free energy $F$. The following argument shows that $Z = \exp(-F/k_B T)$:

$$S = -k_B \sum_m p_m \ln p_m = -k_B \sum_m p_m \ln\left[\frac{\exp(-\varepsilon_m/k_B T)}{Z(T,N,V)}\right] = (1/T)\sum_m p_m \varepsilon_m + k_B \ln[Z(T,N,V)] \sum_m p_m$$

$$\rightarrow \quad TS = U + k_B T \ln[Z(T,N,V)] \quad \rightarrow \quad Z(T,N,V) = \exp[-(U-TS)/k_B T]. \tag{1}$$

Note that the above arguments are predicated on the assumption that the total number $N$ of particles in the system is constant.

**3. Useful thermodynamic identities**. The following thermodynamic identities are useful when dealing with a system of fixed volume $V$ and fixed number of particles $N$, which is in thermal equilibrium while in thermal contact with a reservoir at temperature $T$. Some of these identities were used by Einstein in his original papers on thermal radiation [5–10].

*Partition function*: $\quad Z(T,V,N) = \sum_m \exp(-\varepsilon_m/k_B T) \quad \rightarrow \quad \frac{\partial Z}{\partial T} = \frac{1}{k_B T^2} \sum_m \varepsilon_m \exp\left(-\frac{\varepsilon_m}{k_B T}\right). \tag{2}$

*Average energy*: $\quad \mathcal{E}(T,V,N) = \sum_m \varepsilon_m \exp(-\varepsilon_m/k_B T)/Z$

$$\rightarrow \quad \frac{\partial \mathcal{E}}{\partial T} = \frac{1}{k_B T^2} \sum_m \varepsilon_m^2 \exp\left(-\frac{\varepsilon_m}{k_B T}\right)/Z - \frac{(\partial Z/\partial T)}{Z^2} \sum_m \varepsilon_m \exp\left(-\frac{\varepsilon_m}{k_B T}\right)$$

$$\rightarrow \quad \frac{\partial \mathcal{E}}{\partial T} = \frac{\langle \varepsilon^2 \rangle - \langle \varepsilon \rangle^2}{k_B T^2} = \frac{\text{Var}(\mathcal{E})}{k_B T^2}. \tag{3}$$

*Entropy*: $\quad S(T,V,N) = -k_B \sum_m p_m \ln p_m$

$$= k_B \sum_m Z^{-1} \exp(-\varepsilon_m/k_B T)[\varepsilon_m/(k_B T) + \ln Z]$$

$$= (\mathcal{E}/T) + k_B \ln Z. \tag{4}$$

$$\rightarrow \quad \frac{\partial S}{\partial T} = \frac{\partial \mathcal{E}/\partial T}{T} - \frac{\mathcal{E}}{T^2} + k_B (\partial Z/\partial T)/Z = \frac{\partial \mathcal{E}/\partial T}{T} = \frac{\text{Var}(\mathcal{E})}{k_B T^3}. \tag{5}$$

$$\rightarrow \quad \frac{\partial S}{\partial \mathcal{E}} = \frac{\partial S/\partial T}{\partial \mathcal{E}/\partial T} = \frac{1}{T}. \tag{6}$$

$$\rightarrow \quad \frac{\partial^2 S}{\partial \mathcal{E}^2} = \frac{\partial}{\partial \mathcal{E}}\left(\frac{\partial S}{\partial \mathcal{E}}\right) = \frac{\partial}{\partial \mathcal{E}}\left(\frac{1}{T}\right) = -\frac{\partial T/\partial \mathcal{E}}{T^2} = -\frac{1}{T^2(\partial \mathcal{E}/\partial T)} = -\frac{k_B}{\text{Var}(\mathcal{E})}$$

$$\rightarrow \quad \text{Var}(\mathcal{E}) = -k_B \left(\frac{\partial^2 S}{\partial \mathcal{E}^2}\right)^{-1}. \tag{7}$$

In the case of a photon gas trapped within (and in equilibrium with the walls of) an otherwise empty chamber at temperature $T$, the above identities may be verified as follows (see Ref.[1], Eq.(35)):

$$\mathcal{E}(T,V) = \left(\frac{\pi^2 k_B^4}{15 c^3 \hbar^3}\right) V T^4 \quad \rightarrow \quad \text{Var}(\mathcal{E}) = k_B T^2 \left(\frac{\partial \mathcal{E}}{\partial T}\right) = \left(\frac{4\pi^2 k_B^5}{15 c^3 \hbar^3}\right) V T^5. \tag{8}$$

$$S(T,V) = \left(\frac{4\pi^2 k_B^4}{45 c^3 \hbar^3}\right) V T^3 \quad \rightarrow \quad \text{Var}(\mathcal{E}) = k_B T^3 \left(\frac{\partial S}{\partial T}\right) = \left(\frac{4\pi^2 k_B^5}{15 c^3 \hbar^3}\right) V T^5. \tag{9}$$

$$S(\mathcal{E},V) = \tfrac{4}{3}\left(\sqrt[4]{\tfrac{\pi^2 k_B^4 V}{15 c^3 \hbar^3}}\right)\mathcal{E}^{3/4} \quad \rightarrow \quad \frac{\partial S}{\partial \mathcal{E}} = \left(\sqrt[4]{\tfrac{\pi^2 k_B^4 V}{15 c^3 \hbar^3}}\right)\mathcal{E}^{-1/4} = \frac{1}{T}. \tag{10}$$

$$\rightarrow \quad \frac{\partial^2 S}{\partial \mathcal{E}^2} = -\tfrac{1}{4}\left(\sqrt[4]{\tfrac{\pi^2 k_B^4 V}{15 c^3 \hbar^3}}\right)\mathcal{E}^{-5/4} = -\left(\tfrac{4\pi^2 k_B^4 V T^5}{15 c^3 \hbar^3}\right)^{-1}$$

$$\rightarrow \quad \mathrm{Var}(\mathcal{E}) = -k_B\left(\tfrac{\partial^2 S}{\partial \mathcal{E}^2}\right)^{-1} = \left(\tfrac{4\pi^2 k_B^5}{15 c^3 \hbar^3}\right)VT^5. \tag{11}$$

**4. The peak frequency of blackbody radiation**. The radiant energy per unit area per unit time per unit angular frequency emanating from a blackbody at temperature $T$ was given in Ref.[1], Eq.(10) as follows:

$$I(\omega, T) = \frac{\hbar \omega^3}{4\pi^2 c^2 [\exp(\hbar \omega / k_B T) - 1]}. \tag{12}$$

Defining $x = \hbar \omega / k_B T$, the above expression may be written $[(k_B T)^3/(2\pi c \hbar)^2] x^3/[\exp(x) - 1]$. The derivative of this expression with respect to $x$ vanishes when $\exp(-x) = 1 - (x/3)$. The solutions of this equation, which can be found graphically, are $x = 0$ and $x \cong 2.82144$. The peak intensity thus occurs at $\omega_\mathrm{peak} \cong 2.82144\, k_B T/\hbar$, where $I(\omega_\mathrm{peak}, T) \cong 1.4214 (k_B T)^3/(2\pi c \hbar)^2$.

**5. Blackbody radiation fluctuations in the classical limit when $\hbar$ goes to zero**. According to the Rayleigh-Jeans theory, each EM mode inside a blackbody cavity has $k_B T$ of energy (on average), which is divided equally between its constituent electric and magnetic field components. The Maxwell-Lorentz electrodynamics estimates the number of modes per unit volume in the frequency range $(\omega, \omega + \mathrm{d}\omega)$ to be $(\omega^2/\pi^2 c^3)\mathrm{d}\omega$. Let us denote by $E_0$ the $E$-field amplitude of a single mode having frequency $\omega$ in thermal equilibrium with the cavity walls. The total energy of the mode integrated over the volume $V$ of the cavity is thus given by $\mathbb{E} = \tfrac{1}{2}\varepsilon_0 E_0^2 V$. In general, the single-mode energy $\mathbb{E}$ is a function of the temperature $T$, which could also depend on frequency $\omega$ (e.g., the Planck mode), but it does not depend on the propagation direction $\boldsymbol{k}$ inside the cavity, nor does it depend on the polarization state of the mode. Most significantly, the mode energy $\mathbb{E}$ does not depend on the volume $V$ of the cavity. If the cavity becomes larger, the $E$-field amplitude $E_0$ must become correspondingly smaller, so that the mode energy $\mathbb{E}$ would remain constant.

Consider a small volume $v \ll \lambda^3$, where $\lambda = 2\pi c/\omega$ is the wavelength associated with the frequency $\omega$ of the mode. Given that (on average) the total energy of the modes in the frequency range $(\omega, \omega + \mathrm{d}\omega)$ throughout the entire volume $V$ at any given instant of time will be $(\omega^2 V \mathbb{E}/\pi^2 c^3)\mathrm{d}\omega$, multiplication by $v/V$ yields the corresponding (average) energy $\tilde{E}$ within the small volume $v$ as follows:

$$\mathrm{Ave}[\tilde{E}(\omega, T, v)] = (\omega^2 v / \pi^2 c^3)(\tfrac{1}{2}\varepsilon_0 E_0^2 V)\mathrm{d}\omega = \left(\tfrac{\omega^2 \mathbb{E}}{\pi^2 c^3}\right) v\, \mathrm{d}\omega = \mathcal{E}(\omega, T) v\, \mathrm{d}\omega. \tag{13}$$

Next, we examine energy fluctuations within a small volume $v$. Taking a single mode of energy $\mathbb{E}$ as a micro-canonical ensemble, statistical mechanics requires that $\mathrm{Var}(\mathbb{E}) = k_B T^2(\partial \mathbb{E}/\partial T)$; see Eq.(3). In the Rayleigh-Jeans regime, where $\mathbb{E} = k_B T$, we find $\mathrm{Var}(\mathbb{E}) = (k_B T)^2 = \mathbb{E}^2$. Since the various modes are independent of each other, the total variance of energy within the volume $V$ of the cavity in the frequency range $(\omega, \omega + \mathrm{d}\omega)$ is readily seen to be $(\omega^2 V \mathbb{E}^2/\pi^2 c^3)\mathrm{d}\omega$. If we now assume that, within non-overlapping small volumes $v$, energy fluctuations occur independently of each other, we must multiply the above total variance by $v/V$ to arrive at

$$\mathrm{Var}[\tilde{E}(\omega, T, v)] = (\omega^2 \mathbb{E}^2 v/\pi^2 c^3)\mathrm{d}\omega = \pi^2 c^3 [\mathcal{E}(\omega, T)/\omega]^2 v\, \mathrm{d}\omega. \tag{14}$$

A comparison with Eq.(33) in Ref.[1] now reveals that these classical fluctuations are consistent with Planck's blackbody radiation formula in the limit when $\hbar \to 0$.

**6. Blackbody radiation fluctuations traceable to the discrete nature of photons.** Consider $N$ identical, non-interacting particles distributed randomly and independently within an otherwise empty box of volume $V$. The probability that $n$ of these particles end up together in a smaller volume $v$ inside the box is

$$p_n = \binom{N}{n} \left(\frac{v}{V}\right)^n \left(1 - \frac{v}{V}\right)^{N-n}. \tag{15}$$

The binomial expansion formula $(x+y)^N = \sum_{n=0}^{N} \binom{N}{n} x^n y^{N-n}$ confirms that $\sum_{n=0}^{N} p_n = 1$. Setting $y = 1 - x$, we proceed to compute the 1st and 2nd derivatives with respect to $x$ of the following identity:

$$S(x) = \sum_{n=0}^{N} \binom{N}{n} x^n (1-x)^{N-n} = 1. \tag{16}$$

1st derivative: $S'(x) = \sum \binom{N}{n} [n x^{n-1}(1-x)^{N-n} - (N-n) x^n (1-x)^{N-n-1}] = 0$

$$\to \quad \left(\tfrac{1}{x} + \tfrac{1}{1-x}\right) \sum n \binom{N}{n} x^n (1-x)^{N-n} - \tfrac{N}{1-x} \sum \binom{N}{n} x^n (1-x)^{N-n} = 0$$

$$\to \quad \sum_{n=0}^{N} n \binom{N}{n} x^n (1-x)^{N-n} = Nx. \tag{17}$$

A comparison of Eq.(15) with Eq.(17) reveals that $\langle n \rangle = Nv/V$. As for the variance of $n$, we need to differentiate Eq.(17) with respect to $x$. We will have

2nd derivative: $\sum n \binom{N}{n} [n x^{n-1}(1-x)^{N-n} - (N-n) x^n (1-x)^{N-n-1}] = N$

$$\to \quad \left(\tfrac{1}{x} + \tfrac{1}{1-x}\right) \sum n^2 \binom{N}{n} x^n (1-x)^{N-n} - \tfrac{N}{1-x} \sum n \binom{N}{n} x^n (1-x)^{N-n} = N$$

$$\to \quad \sum n^2 \binom{N}{n} x^n (1-x)^{N-n} = N^2 x^2 + Nx(1-x). \tag{18}$$

Comparing Eq.(15) with Eq.(18), we find the variance of $n$, as follows:

$$\text{Var}(n) = \langle n^2 \rangle - \langle n \rangle^2 = N(v/V) - N(v/V)^2 = [1 - (v/V)]\langle n \rangle. \tag{19}$$

If $v \ll V$, then $\text{Var}(n) \cong \langle n \rangle$. Suppose now that the particles are known to have a fixed energy, say, they are identical photons of energy $\hbar\omega$. Then $\langle \mathcal{E} \rangle = \langle n \rangle \hbar\omega$ while $\text{Var}(\mathcal{E}) \cong \langle n \rangle (\hbar\omega)^2 = \hbar\omega \langle \mathcal{E} \rangle$. As pointed out by Einstein [6,7], this constitutes the first term in the variance of the energy-density of blackbody radiation appearing on the right-hand-side of Eq.(33) in Ref. [1].

In contrast, if the chamber is filled with a simple (monatomic) gas, then the average energy of the gas occupying volume $v$ will be $\langle \mathcal{E} \rangle = 3\langle n \rangle k_B T/2$, but computing the variance of energy will require accounting for fluctuations in the energy of each particle in addition to those of the number $n$ of particles.

**7. Johnson noise and the Nyquist theorem.** In the $RLC$ circuit shown in the figure, the $n$-turn inductor has length $\ell$, cross-sectional area $\mathcal{A}$, and inductance $L = \Phi_0/I_0 = n\mathcal{A}\mu_0 H/I_0 = \mu_0 n^2 \mathcal{A}/\ell$. Here $\Phi_0$ is the total magnetic flux, and $H = B/\mu_0 = nI_0/\ell$ is the magnetic field inside the inductor—both produced by the current $I_0$. Assuming the thermal noise that accompanies the resistance $R$ is modelled as a separate voltage source, $v_s(t)$, and denoting the resulting current by $i(t)$, the current $I(\omega)\exp(-i\omega t)$ and voltage $V_s(\omega)\exp(-i\omega t)$ at the frequency $\omega$ will be related by

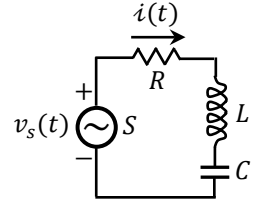

$$V_s(\omega) = \left[R + i\left(L\omega - \frac{1}{C\omega}\right)\right]I(\omega). \tag{20}$$

At the resonance frequency $\omega_0$, the impedance of the circuit is at a minimum and, therefore, $LC\omega_0^2 = 1$. If the circuit happens to have a large quality-factor $Q$, the excited frequencies will be confined to a narrow band centered at $\omega_0$. The magnetic field energy stored in the inductor is the product of the $H$-field energy-density and the volume of the inductor, that is,

$$\mathcal{E}(\omega) = \tfrac{1}{2}\mu_0 H^2 \mathcal{A}\ell = \tfrac{1}{2}\mu_0 (nI/\ell)^2 \mathcal{A}\ell = \tfrac{1}{2}L|I(\omega)|^2. \tag{21}$$

Considering that the current is sinusoidal, namely, $|I(\omega)|\sin(\omega t + \varphi)$, the time-averaged energy stored in the inductor is only one-half of $\mathcal{E}(\omega)$ given by Eq.(21). However, since we are interested in the total energy of the system (that is, the sum of the energies stored in the inductor and capacitor), the time-independent $\mathcal{E}(\omega)$ of Eq.(21) is a good estimate for the *total* energy. From Eq.(20) we find

$$|I(\omega)|^2 = \frac{V_s^2(\omega)}{R^2 + \left(L\omega - \frac{1}{C\omega}\right)^2} = \frac{V_s^2(\omega_0 + \delta\omega)}{R^2 + \left[L(\omega_0 + \delta\omega) - \frac{1}{C(\omega_0 + \delta\omega)}\right]^2} \cong \frac{V_s^2(\omega_0 + \delta\omega)}{R^2 + \left[L(\omega_0 + \delta\omega) - \frac{\omega_0 - \delta\omega}{C\omega_0^2}\right]^2} = \frac{V_s^2(\omega_0 + \delta\omega)}{R^2 + 4L^2(\delta\omega)^2}. \tag{22}$$

Since the assumed $RLC$ circuit has a narrow bandwidth, it is reasonable to suppose that $V_s(\omega)$ over this bandwidth is fairly uniform, in which case, the integral of the squared current becomes

$$\int_{\omega=0}^{\infty} |I(\omega)|^2 \mathrm{d}\omega \cong \int_{-\infty}^{\infty} \frac{V_s^2(\omega_0 + \zeta)}{R^2 + 4L^2\zeta^2}\mathrm{d}\zeta \cong \frac{V_s^2(\omega_0)}{R^2}\int_{-\infty}^{\infty} \frac{\mathrm{d}\zeta}{1 + (2L\zeta/R)^2} = \frac{V_s^2(\omega_0)}{2RL}\int_{-\infty}^{\infty} \frac{\mathrm{d}x}{1+x^2} = \frac{\pi V_s^2(\omega_0)}{2RL}. \tag{23}$$

The average electromagnetic energy stored in the form of magnetic field in the inductance $L$ and also in the form of electric field in the capacitance $C$ is thus given by Eqs.(21) and (23), as follows:

$$\langle \mathcal{E}(\omega_0) \rangle = \tfrac{1}{2}L \int_0^{\infty} |I(\omega)|^2 \mathrm{d}\omega = \frac{\pi V_s^2(\omega_0)}{4R} = \tfrac{1}{2}k_B T. \tag{24}$$

In the above equation, we have invoked the equipartition theorem of statistical mechanics [2−4], which states that each degree of freedom of a system in thermal equilibrium at temperature $T$ has, on average, the energy $\tfrac{1}{2}k_B T$, where $k_B$ is the Boltzmann constant. The noise power-density per unit (angular) frequency associated with the resistance $R$ is thus given by

$$V_s^2(\omega)/R = (2/\pi)k_B T. \tag{25}$$

Note that the noise power-density per unit frequency is independent of the frequency $\omega$; this is characteristic of white noise. The noise power within a frequency range $\mathrm{d}\omega = 2\pi \mathrm{d}f$ is thus given by $P(\omega)\mathrm{d}\omega = (2/\pi)k_B T \mathrm{d}\omega = 4k_B T \mathrm{d}f$. This result is often referred to as the Johnson-Nyquist theorem.

**8. An alternative derivation of the Johnson-Nyquist theorem**. Let the resistor $R$ be connected to another resistor having the same resistance via a perfectly matched transmission line. The length of the transmission line is $\ell$, its characteristic impedance is $R$, and the propagation velocity of the EM fields along the line is $V$. If we model the noise produced by the resistor at temperature $T$ as a voltage source $v_s(t)$ connected in series to the resistor, then the current flowing in the circuit will be $i(t) = v_s(t)/2R$, which transmits an average electrical power $\langle Ri^2(t) \rangle = \langle v_s^2(t) \rangle/4R$ to the second resistor. Here the angled brackets represent statistical averaging.

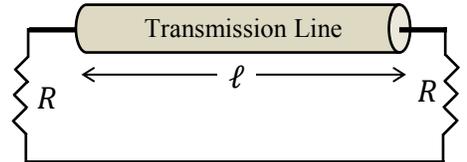

The EM fields inside the transmission line may now be considered as blackbody radiation in a one-dimensional box. The allowed frequencies $\omega_n$ inside the box must have wavelengths $\lambda_n$ such that the length $\ell$ of the box is an integer multiple of $\lambda_n/2$, that is, $\ell = n\lambda_n/2 = n\pi V/\omega_n$. Consequently, $\omega_n = n\pi V/\ell$. The allowed frequencies are thus seen to be separated from each other by $\delta\omega = \pi V/\ell$, indicating that, within a range of frequencies $d\omega = 2\pi df$, the number of allowed modes inside the transmission line is $d\omega/\delta\omega = (2\ell/V)df$.

In accordance with the Planck distribution, the average energy content of each mode is given by $\langle \mathcal{E} \rangle = \hbar\omega/[\exp(\hbar\omega/k_B T) - 1]$, which, in the classical limit (i.e., $\hbar\omega \ll k_B T$), reduces to $\langle \mathcal{E} \rangle \cong k_B T$. Needless to say, this is twice the thermal energy content per mode, $\frac{1}{2}k_B T$, associated with a classical oscillator in one dimension. The factor of 2 arises from the fact that each mode of the transmission line could have two independent polarization states. All in all, the blackbody radiation inside the transmission line has energy $(2k_B T\ell/V)df$ within the frequency range $df$. Dividing this entity by $\ell$ yields the energy per unit length, a further division by 2 yields the energy density that flows either to the right or the left along the transmission line, and, finally, multiplication by the wave velocity $V$ yields the rate at which EM energy is transferred to each resistor at either end of the line, namely, $k_B T df$. A comparison with our earlier result now shows that $\langle v_s^2(t) \rangle = 4k_B TR df$. This completes our alternative proof of the Johnson-Nyquist theorem.

**9. Photodetection statistics**. The following argument follows largely in the footsteps of E.M. Purcell [11]. Let the cycle-averaged (also known as low-pass-filtered) intensity of quasi-monochromatic radiation arriving at an ideal, small-area photodetector be denoted by $I(t)$. Within a short time-interval $(t, t + \Delta t)$, the probability that a single photo-electron is released is $\alpha I(t)\Delta t$, where $\alpha$ is a characteristic parameter of the photodetector for the incident radiation, which is centered at the frequency $\omega_0$ and has linewidth $\Delta\omega$. We denote by $p(n,t)$ the probability that $n$ photo-electrons are released in the time interval $[0,t]$. For $n = 0$ we will have

$$p(0, t + \Delta t) = p(0,t)[1 - \alpha I(t)\Delta t] \quad \rightarrow \quad \frac{dp(0,t)}{dt} = -\alpha I(t)p(0,t)$$

$$\rightarrow \quad p(0,t) = \exp\left[-\alpha \int_0^t I(t')dt'\right]. \tag{26}$$

Similarly, the probability that $n$ photo-electrons are released between $t = 0$ and $t$ can be derived from the following first-order difference-differential equation:

$$p(n+1, t+\Delta t) = p(n+1,t)[1 - \alpha I(t)\Delta t] + p(n,t)\alpha I(t)\Delta t$$

$$\rightarrow \quad \frac{dp(n+1,t)}{dt} = -\alpha I(t)[p(n+1,t) - p(n,t)]. \tag{27}$$

Using the method of proof by induction and with the aid of Eq.(26), we arrive at

$$p(n,t) = \frac{1}{n!}\left[\alpha \int_0^t I(t')dt'\right]^n \exp\left[-\alpha \int_0^t I(t')dt'\right]. \tag{28}$$

The probability distribution function in Eq.(28) is known as the Poisson distribution. Note that, for $t \geq 0$, we have $p(n,t) \geq 0$ for $n = 0, 1, 2, \cdots$, and that $\sum_{n=0}^{\infty} p(n,t) = 1$, as expected from a probability distribution function. In a photon-counting experiment during the fixed time interval $[0,T]$, we may define the parameter $\eta = \alpha \int_0^T I(t)dt$ and write $p(n,T) = \exp(-\eta)\eta^n/n!$. The average number $\langle n \rangle$ of counted photo-electrons can then be computed as follows:

$$\langle n \rangle = \sum_{n=0}^{\infty} np(n,T) = \exp(-\eta)\sum_{n=0}^{\infty} n\eta^n/n! = \eta\exp(-\eta)\sum_{n=1}^{\infty} \eta^{n-1}/(n-1)! = \eta. \tag{29}$$

Similarly, $\langle n^2 \rangle$, the average squared number of photo-electrons, is found to be

$$\langle n^2 \rangle = \sum_{n=0}^{\infty} n^2 p(n,T) = \exp(-\eta)\sum_{n=0}^{\infty} n^2\eta^n/n! = \exp(-\eta)\sum_{n=1}^{\infty} n\eta^n/(n-1)!$$

$$= \exp(-\eta)\sum_{n=1}^{\infty} (n-1+1)\eta^n/(n-1)!$$

$$= \exp(-\eta)\,[\sum_{n=2}^{\infty} \eta^n/(n-2)! + \sum_{n=1}^{\infty} \eta^n/(n-1)!]$$

$$= \exp(-\eta)\,[\eta^2 \sum_{n=0}^{\infty}(\eta^n/n!) + \eta \sum_{n=0}^{\infty}(\eta^n/n!)] = \eta^2 + \eta. \tag{30}$$

The variance of the number of photo-electrons released in the interval $[0,T]$ is thus given by

$$\mathrm{Var}(n) = \mathrm{Ave}(n - \langle n \rangle)^2 = \langle n^2 \rangle - \langle n \rangle^2 = \eta. \tag{31}$$

So far, our assumption has been that the incident intensity $I(t)$ is a predetermined function of time. If $I(t)$ happens to be a stationary random process having ensemble averages $\langle I(t) \rangle = I_o$ and $\langle I(t)I(t+\tau)\rangle/\langle I(t)\rangle^2 = g^{(2)}(\tau)$, then the ensemble averages of $\eta$ and $\eta^2$ will become

$$\langle \eta \rangle = \alpha \int_0^T \langle I(t) \rangle \mathrm{d}t = \alpha I_o T. \tag{32}$$

$$\langle \eta^2 \rangle = \alpha^2 \langle \int_{t=0}^T \int_{t'=0}^T I(t)I(t')\mathrm{d}t\mathrm{d}t' \rangle$$

$$= \alpha^2 \langle \int_{\tau=0}^T \int_{t=0}^{T-\tau} I(t)I(t+\tau)\mathrm{d}t\mathrm{d}\tau + \int_{\tau=0}^T \int_{t=\tau}^T I(t)I(t-\tau)\mathrm{d}t\mathrm{d}\tau \rangle$$

$$= \alpha^2 \langle I(t) \rangle^2 \int_{\tau=-T}^T (T - |\tau|) g^{(2)}(\tau) \mathrm{d}\tau$$

$$= \alpha^2 I_o^2\, T \int_{-T}^T [1 - (|\tau|/T)] g^{(2)}(\tau) \mathrm{d}\tau$$

> The width of $g^{(2)}(\tau) - 1$, which is inversely proportional to $\Delta\omega$, is assumed to be $\ll T$.

$$\cong (\alpha I_o T)^2 \left\{ 1 + T^{-1} \int_{-T}^T [g^{(2)}(\tau) - 1] \mathrm{d}\tau \right\}. \tag{33}$$

The ensemble-averaged variance of the number of photo-electrons released in $[0,T]$ is given by

$$\mathrm{Var}(n) = \langle n^2 \rangle - \langle n \rangle^2 = \langle \eta^2 \rangle + \langle \eta \rangle - \langle \eta \rangle^2 = \alpha I_o T \left\{ 1 + \alpha I_o \int_{-T}^T [g^{(2)}(\tau) - 1] \mathrm{d}\tau \right\}. \tag{34}$$

Noting that $\langle n \rangle = \alpha I_o T$ and denoting the area under $g^{(2)}(\tau) - 1$ by $\tau_o$, Eq.(34) may be written as

$$\mathrm{Var}(n) = \langle n \rangle [1 + (\langle n \rangle \tau_o/T)]. \tag{35}$$

If the incident light happens to be fully unpolarized, then the two orthogonal components of polarization, say, $p$ and $s$, will have similar intensities while remaining uncorrelated. They will thus give rise to equal average photo-electron numbers, $\langle n_p \rangle = \langle n_s \rangle$, and also equal variances of the photo-electron numbers, $\mathrm{Var}(n_p) = \mathrm{Var}(n_s)$. One can then use Eq.(35) to write

$$\mathrm{Var}(n_{\mathrm{total}}) = \mathrm{Var}(n_p) + \mathrm{Var}(n_s) = \langle n_p \rangle [1 + (\langle n_p \rangle \tau_o/T)] + \langle n_s \rangle [1 + (\langle n_s \rangle \tau_o/T)]$$

$$= \langle n_{\mathrm{total}} \rangle [1 + \tfrac{1}{2}(\langle n_{\mathrm{total}} \rangle \tau_o/T)]. \tag{36}$$

In a somewhat modified experiment, one may use a 50/50 beam-splitter to divide a linearly polarized, quasi-monochromatic incident beam between two identical small-area photodetectors. For these two detector, $\langle n_1 \rangle = \langle n_2 \rangle$ and $\mathrm{Var}(n_1) = \mathrm{Var}(n_2)$, where the individual averages and variances satisfy Eq.(35). If we now add the two photo-electron counts together to form $n = n_1 + n_2$, we will have $\langle n \rangle = \langle n_1 \rangle + \langle n_2 \rangle$ and $\langle n^2 \rangle = \langle n_1^2 \rangle + \langle n_2^2 \rangle + 2\langle n_1 n_2 \rangle$. Consequently, $\mathrm{Var}(n) = \mathrm{Var}(n_1) + \mathrm{Var}(n_2) + 2(\langle n_1 n_2 \rangle - \langle n_1 \rangle \langle n_2 \rangle)$. Substitution into Eq.(35) now yields

$$\langle n_1 n_2 \rangle = \langle n_1 \rangle \langle n_2 \rangle [1 + (\tau_o/T)]. \tag{37}$$

This positive cross-correlation between the photo-electron counts of the two detectors is an alternative expression of the result obtained in the Hanbury Brown-Twiss experiment [12].

For a single-mode, quasi-monochromatic laser beam, whose cycle-averaged intensity is nearly constant, we will have $g^{(2)}(\tau) \cong 1$, resulting in pure Poisson statistics for the number of released photo-electrons, where $\text{Var}(n) = \langle n \rangle = \alpha I_0 T$. In this case the excess noise, i.e., the second term on the right-hand-side of Eq.(35), is absent.

In the case of blackbody radiation (at fixed temperature $\tilde{T}$) from a small aperture of area $A$, center frequency $\omega$, linewidth $\Delta\omega$, and integration time $T$, Eq.(32) of Ref.[1] gives the single-mode photon number variance as $\text{Var}(m) = \exp(\hbar\omega/k_B\tilde{T})/[\exp(\hbar\omega/k_B\tilde{T}) - 1]^2 = \langle m \rangle + \langle m \rangle^2$. This variance must be multiplied by the density (per unit volume) of modes, $\omega^2\Delta\omega/(\pi^2 c^3)$, by the relevant mode volume $cAT$, by a factor ½ to account for the fraction of photons that exit the blackbody through the aperture, and by another factor of ½ to account for the average obliquity factor $\langle\cos\theta\rangle$. The first term in the expression of the photon-number variance, $A(\omega/2\pi c)^2\langle m\rangle T\Delta\omega$, thus corresponds to the first term $\langle n_{\text{total}}\rangle = \alpha I_0 T$ on the right-hand-side of Eq.(36). Noting that $\lambda = 2\pi c/\omega$, we may write $\alpha I_0 = (A/\lambda^2)\langle m\rangle\Delta\omega$. The second term in the expression of the photon-number variance for blackbody radiation then becomes $(A/\lambda^2)\langle m\rangle^2 T\Delta\omega$. Agreement with the second term on the right-hand-side of Eq.(36) above requires that $A \sim \lambda^2$ and $\tau_0 = \int_{-T}^{T}[g^{(2)}(\tau) - 1]d\tau \sim (\Delta\omega)^{-1}$. The first requirement is satisfied because the detector must act as a point-detector for the blackbody radiation emerging from the small aperture. The second requirement is generally satisfied for chaotic (or thermal) light [13]. The excess noise of blackbody radiation thus complies with Eq.(36).

**10. States of a single particle in a rigid box**. Consider a rectangular parallelepiped box of dimensions $L_x \times L_y \times L_z$. The evacuated box, held at a constant temperature $T$, contains a single point-particle of mass $m$, velocity $\boldsymbol{v}$, momentum $\boldsymbol{p} = m\boldsymbol{v}$, and kinetic energy $\mathcal{E} = \tfrac{1}{2}mv^2 = p^2/(2m)$. A plane-wave solution of Schrödinger's equation $-(\hbar^2/2m)\nabla^2\psi(\boldsymbol{r},t) = i\hbar\,\partial\psi(\boldsymbol{r},t)/\partial t$ in free space is given by $\psi(\boldsymbol{r},t) = \psi_0 \exp[i(\boldsymbol{k}\cdot\boldsymbol{r} - \omega t)]$, where $\boldsymbol{k} = \boldsymbol{p}/\hbar$ and $\omega = \mathcal{E}/\hbar$, with $\hbar$ being Planck's reduced constant. To ensure that the wave-function vanishes at the interior walls of the box, we need a superposition of eight plane-waves having $k$-vectors $\boldsymbol{k} = \pm(\pi n_x/L_x)\hat{\boldsymbol{x}} \pm (\pi n_y/L_y)\hat{\boldsymbol{y}} \pm (\pi n_z/L_z)\hat{\boldsymbol{z}}$, where $n_x, n_y, n_z$ are positive integers $1, 2, 3, \cdots$. The combined wave-function will be

$$\begin{aligned}\psi(\boldsymbol{r},t) &= \psi_0\{\exp[i(k_x x + k_y y + k_z z)] - \exp[i(k_x x + k_y y - k_z z)] - \exp[i(k_x x - k_y y + k_z z)] \\ &\quad + \exp[i(k_x x - k_y y - k_z z)] - \exp[i(-k_x x + k_y y + k_z z)] + \exp[i(-k_x x + k_y y - k_z z)] \\ &\quad + \exp[i(-k_x x - k_y y + k_z z)] - \exp[i(-k_x x - k_y y - k_z z)]\}\exp(-i\omega t) \\ &= 2i\psi_0\{\exp[i(k_x x + k_y y)] - \exp[i(k_x x - k_y y)] - \exp[i(-k_x x + k_y y)] + \exp[i(-k_x x - k_y y)]\} \\ &\quad \times \sin(k_z z)\exp(-i\omega t) \\ &= -4\psi_0[\exp(ik_x x) - \exp(-ik_x x)]\sin(k_y y)\sin(k_z z)\exp(-i\omega t) \\ &= -8i\psi_0 \sin(k_x x)\sin(k_y y)\sin(k_z z)\exp(-i\omega t). \end{aligned} \quad (38)$$

The amplitude $\psi_0$ may be determined (aside from an arbitrary phase-factor) by requiring the integral of $|\psi(\boldsymbol{r},t)|^2$ over the volume of the box to be unity. Defining $V = L_x L_y L_z$, we will have $|\psi_0| = 1/\sqrt{8V}$. The desired wave-function is thus given by

$$\psi(\boldsymbol{r},t) = 2\sqrt{2/V}\sin(k_x x)\sin(k_y y)\sin(k_z z)\exp[-i(\omega t - \varphi_0)]. \quad (39)$$

The above wave-function with $(k_x, k_y, k_z) = (\pi n_x/L_x, \pi n_y/L_y, \pi n_z/L_z)$, where $(n_x, n_y, n_z)$ are positive integers, represents all possible states of a single point-particle inside a rigid box of dimensions $L_x \times L_y \times L_z$ and volume $V$. The particle's linear momentum and kinetic energy are readily found to be

$$\boldsymbol{p} = \hbar\boldsymbol{k} = \pi\hbar(n_x/L_x, n_y/L_y, n_z/L_z). \quad (40)$$

$$\mathcal{E} = \hbar\omega = \hbar^2 k^2/2m = (\pi^2\hbar^2/2m)\left[(n_x/L_x)^2 + (n_y/L_y)^2 + (n_z/L_z)^2\right]. \tag{41}$$

**11. Partition function for particles in a box and derivation of the Sackur-Tetrode equation.** For the single particle in a box described in the preceding section, the partition function is the sum over all possible states $\ell$ of the exponential factor $\exp(-\mathcal{E}_\ell/k_B T)$. We will have

$$Z(T, N=1, V) = \sum_{n_x, n_y, n_z} \exp\{-(\pi^2\hbar^2/2mk_B T)[(n_x/L_x)^2 + (n_y/L_y)^2 + (n_z/L_z)^2]\}$$

$$\cong \int_{n_x=0}^{\infty}\int_{n_y=0}^{\infty}\int_{n_z=0}^{\infty} \exp\{-(\pi^2\hbar^2/2mk_B T)[(n_x/L_x)^2 + (n_y/L_y)^2 + (n_z/L_z)^2]\}\,\mathrm{d}n_x \mathrm{d}n_y \mathrm{d}n_z$$

$$= (2mk_B T/\pi^2\hbar^2)^{3/2} L_x L_y L_z \iiint_0^\infty \exp[-(x^2+y^2+z^2)]\,\mathrm{d}x\mathrm{d}y\mathrm{d}z \quad \leftarrow \boxed{\int_0^\infty \exp(-x^2)\,\mathrm{d}x = \sqrt{\pi}/2}$$

$$= (mk_B T/2\pi\hbar^2)^{3/2} V. \tag{42}$$

For a system of $N$ identical, non-interacting particles, $Z(T,N,V) = Z(T,1,V)^N/N!$. This is because $\sum_{\ell_1}\sum_{\ell_2}\cdots\sum_{\ell_N} \exp[-(\mathcal{E}_{\ell_1} + \mathcal{E}_{\ell_2} + \cdots + \mathcal{E}_{\ell_N})/k_B T] = [\sum_\ell \exp(-\mathcal{E}_\ell/k_B T)]^N$, and also because the particles are indistinguishable, which requires division by $N!$ in order to eliminate the contribution of particle arrangements that are mere permutations of each other. The Helmholtz free energy of the $N$-particle system (described in Section 2) is thus given by

$$F = -k_B T \ln Z(T,N,V) \cong -k_B T \ln\left[(mk_B T/2\pi\hbar^2)^{3N/2} V^N/N!\right] \quad \leftarrow \boxed{\text{Stirling's approximation: } N! \cong N^N e^{-N}\sqrt{N}e}$$

$$\cong -k_B T \ln\left[(mk_B T/2\pi\hbar^2)^{3N/2} V^N/(N^N e^{-N}\sqrt{N}e)\right] \quad \boxed{\text{Small compared to other terms}}$$

$$= -k_B T\left\{N \ln\left[(mk_B T/2\pi\hbar^2)^{3/2}(V/N)\right] + N - \tfrac{1}{2}\ln N - 1\right\}. \tag{43}$$

Since $F = U - TS$, and since, for the ideal gas, $U = 3Nk_B T/2$, the above equation may be written

$$S(T,N,V) \cong Nk_B\left\{\ln\left[(2\pi mk_B T/h^2)^{3/2}(V/N)\right] + \tfrac{5}{2}\right\}. \tag{44}$$

This is the well-known Sackur-Tetrode equation [2], which gives the entropy of an ideal monatomic gas as a function of the temperature $T$, the number of particles $N$, and the volume $V$. Here $k_B = 1.38065 \times 10^{-23}$ Joule/deg is Boltzmann's constant, while $h = 6.626068 \times 10^{-34}$ Joule·sec is the Planck constant. The formula breaks down at low temperatures and/or at high particle densities, partly because of the neglect of the interaction among particles, and partly because of the approximations involved in Eq.(42) in going from a sum to an integral. (The expression for the entropy cannot be valid for very low temperatures, because the entropy does *not* approach a constant value when $T \to 0$ K.) Compare and contrast the entropy of a gas of monatomic particles with the entropy of the photon gas (i.e., blackbody radiation) given in Ref.[1], Eq.(37).

**12. Internal energy, pressure, and chemical potential of the ideal monatomic gas.** One may use the partition function $Z(T,N,V)$ to calculate the internal energy $U$ of the ideal monatomic gas in accordance with Eq.(2), as follows:

$$U = k_B T^2\,\partial \ln[Z(T,N,V)]/\partial T = k_B T^2\,\partial \ln\left[(mk_B T/2\pi\hbar^2)^{3N/2} V^N/N!\right]/\partial T = \tfrac{3}{2}Nk_B T. \tag{45}$$

Alternatively, we can compute the average kinetic energy of a single particle by direct integration over all possible states of the particle, namely,

$$U(T, N = 1, V) = \sum_{n_x,n_y,n_z}(\pi^2\hbar^2/2m)\left[(n_x/L_x)^2 + (n_y/L_y)^2 + (n_z/L_z)^2\right]$$

$$\times \exp\{-(\pi^2\hbar^2/2mk_BT)[(n_x/L_x)^2 + (n_y/L_y)^2 + (n_z/L_z)^2]\}/Z(T, N = 1, V)$$

$$\cong \left[(mk_BT/2\pi\hbar^2)^{3/2}V\right]^{-1}$$

$$\times \int_{n_x=0}^{\infty}\int_{n_y=0}^{\infty}\int_{n_z=0}^{\infty}(\pi^2\hbar^2/2m)[(n_x/L_x)^2 + (n_y/L_y)^2 + (n_z/L_z)^2]$$

$$\times \exp\{-(\pi^2\hbar^2/2mk_BT)[(n_x/L_x)^2 + (n_y/L_y)^2 + (n_z/L_z)^2]\}\,\mathrm{d}n_x\mathrm{d}n_y\mathrm{d}n_z$$

$$= \frac{k_BT(2mk_BT/\pi^2\hbar^2)^{3/2}V}{(mk_BT/2\pi\hbar^2)^{3/2}V}\int_{x=0}^{\infty}\int_{y=0}^{\infty}\int_{z=0}^{\infty}(x^2+y^2+z^2)\exp[-(x^2+y^2+z^2)]\,\mathrm{d}x\mathrm{d}y\mathrm{d}z$$

$$= (8k_BT/\sqrt{\pi^3})\int_0^{\infty}\tfrac{1}{2}\pi r^4\exp(-r^2)\,\mathrm{d}r = \tfrac{3}{2}k_BT. \quad \leftarrow \boxed{\int_0^{\infty}x^4\exp(-x^2)\,\mathrm{d}x = 3\sqrt{\pi}/8} \quad (46)$$

Since the $N$ particles of the system are identical and independent, the total energy of the system will be $U = 3Nk_BT/2$, as before; see Eq.(45). The pressure $p$ on each wall of the container may be derived from the kinetic energies of the individual atoms, $\tfrac{1}{2}m(v_x^2 + v_y^2 + v_z^2)$, by recalling that the component of momentum perpendicular to a given wall, say, $mv_x$, must be multiplied by the number of particles $\tfrac{1}{2}(N/V)v_x$ that strike a unit area of the wall per unit time, then doubled (to account for the change of momentum from before to after collision). We thus find

$$p = \frac{2U}{3V} = \frac{2}{3}u. \tag{47}$$

Here $u = U/V$ is the energy-density of the gas. From Eqs.(46) and (47), one can readily obtain the ideal gas law, namely, $pV = Nk_BT$. Comparing Eq.(47) with Eq.(12) of Ref.[1], we note that the pressure of the ideal monatomic gas as a function of its energy-density is twice that of the photon gas (i.e., blackbody radiation).

The Sackur-Tetrode equation describing the entropy of an ideal monatomic gas consisting of $N$ identical particles of mass $m$ having (absolute) temperature $T$ and volume $V$ is given by Eq.(44). The fundamental thermodynamic identity [2-4] relates the various partial derivatives of the entropy $S(U, N, V)$ to the temperature $T$, chemical potential $\mu_c$, and pressure $p$ of the system, as follows:

$$(\partial S/\partial U)_{N,V} = 1/T; \qquad (\partial S/\partial N)_{U,V} = -\mu_c/T; \qquad (\partial S/\partial V)_{U,N} = p/T. \tag{48}$$

Using Eq.(45), we rewrite the entropy in Eq.(44) as a function of $U$, $N$ and $V$, namely,

$$S(U, N, V) = Nk_B\left\{\ln\left[(4\pi mU/3h^2)^{3/2}(V/N^{5/2})\right] + \tfrac{5}{2}\right\}. \tag{49}$$

In accordance with Eqs.(48), the partial derivatives of the above expression for entropy now yield

$$(\partial S/\partial U)_{N,V} = \tfrac{3}{2}Nk_B/U = 1/T. \tag{50}$$

$$(\partial S/\partial V)_{U,N} = Nk_B/V = p/T. \tag{51}$$

$$\mu_c = -T(\partial S/\partial N)_{U,V} = -k_BT\ln\left[(4\pi mU/3h^2)^{3/2}(V/N^{5/2})\right]$$

$$= -k_BT\ln\left[(2\pi mk_BT/h^2)^{3/2}(V/N)\right]. \tag{52}$$

Considering that the proton mass is $m_p = 1.67262158 \times 10^{-27}$ kg, an order-of-magnitude estimate for the chemical potential given by Eq.(52) is $\mu_c \cong -13k_BT$. This is based on assuming $m = m_p$, $T = 100K$, $V = 1.0\,m^3$, and $N = 6.0221415 \times 10^{23}$, i.e., Avogadro's number. Reducing the temperature by a factor of 10 will change the chemical potential of the gas to $\mu_c \cong -9.55 k_B T$, whereas raising the density $N/V$ by a factor of 10 (while maintaining $T$ at $100K$) will change the chemical potential to $\mu_c \cong -10.7 k_B T$. The more massive the atoms/molecules of the gas are, the more negative its chemical potential becomes.

**Example 1**. Suppose an $N$-particle monatomic gas at temperature $T_1$ and volume $V$ is heated to a new temperature $T_2$ without changing its volume or number of particles. The internal energy of the gas thus increases by $\Delta U = (3/2) N k_B (T_2 - T_1)$, requiring an amount of heat $\Delta Q = \Delta U$. The change in entropy may be calculated as follows:

$$\Delta S = \int_{T_1}^{T_2} \frac{dQ}{T} = \int_{T_1}^{T_2} \frac{(3/2) N k_B dT}{T} = \frac{3}{2} N k_B \ln(T_2/T_1). \tag{53}$$

The above result is seen to be consistent with the Sackur-Tetrode expression for the entropy given by Eq.(44).

**Example 2**. As another example, let an $N$-particle monatomic gas expand reversibly from an initial volume $V_1$ to a final volume $V_2$ while remaining at a constant temperature $T$. The internal energy of the gas will remain at $U = (3/2) N k_B T$, but the pressure $p = N k_B T / V$ will deliver an amount of work $\Delta W$ to the outside world that is given by

$$\Delta W = \int_{V_1}^{V_2} p dV = \int_{V_1}^{V_2} \frac{N k_B T}{V} dV = N k_B T \ln(V_2/V_1). \tag{54}$$

Since the internal energy $U$ of the gas has remained constant, the above work $\Delta W$ must be supplied by the heat bath, which maintains the gas at a fixed temperature. Consequently, $\Delta Q = \Delta W$. The change $\Delta S$ in the entropy of the gas is, therefore, given by

$$\Delta S = \frac{\Delta Q}{T} = \frac{\Delta W}{T} = N k_B \ln(V_2/V_1). \tag{55}$$

Once again, the above result is seen to be consistent with the Sackur-Tetrode expression for the entropy given by Eq.(44).

Note that the entropy, being a property of a system in thermal and diffusional equilibrium, does not depend on the path taken by the system to arrive at its final state. Thus, in a modified version of the present example, shown in the figure below, the entropy of the system will increase by the same amount $\Delta S$ as given by Eq.(55) if the gas were to expand irreversibly and adiabatically from an initial volume $V_1$ to a final volume $V_2$, without adding or subtracting any amount of heat whatsoever. In this system, the diaphragm shatters at some instant of time, thus allowing the gas to expand irreversibly into the evacuated part of the chamber. Since the gas molecules retain their initial kinetic energies, the internal energy $U$ and the temperature $T$ of the gas will not change in consequence of the expansion. The pressure $p$, however, drops in proportion to the ratio $V_1/V_2$, while the entropy increases by $\Delta S = N k_B \ln(V_2/V_1)$.

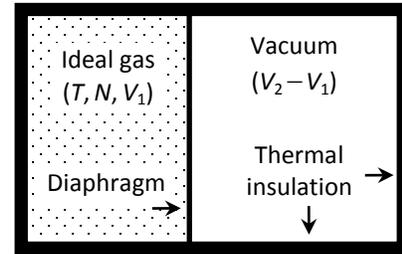

**Example 3**. Suppose a single particle is added to a monatomic $N$-particle gas at temperature $T$ and volume $V$. The particle must have a kinetic energy of $(3/2) k_B T$, which raises the total energy of the system from $U_1 = (3/2) N k_B T$ to $U_2 = (3/2)(N+1) k_B T$. If the goal is to maintain the internal energy of the system at its initial value of $U_1$, then the final temperature must be lowered from $T$ to $NT/(N+1)$. The change of entropy at constant $U$ and $V$ may thus be obtained from Eq.(49), as follows:

$$\Delta S = S(U, N + 1, V) - S(U, N, V) \cong k_B \left\{ \ln[(4\pi mU/3h^2)^{3/2}(V/N^{5/2})] - \frac{5}{2N}^{0} \right\}$$
$$\cong k_B \ln[(2\pi mk_B T/h^2)^{3/2}(V/N)]. \tag{56}$$

According to Eq.(48), the above change in the entropy of the system must be equal to $-\mu_c/T$, in agreement with Eq.(52).

Note that the kinetic energy of the added particle, $(3/2)[N/(N+1)]k_B T$, is essentially the same as that of all the other particles in the system. This energy is very different from the chemical potential $\mu_c$, which has a negative value in the present example. The significance of the chemical potential is in connection with the change in the internal energy, $\Delta U(S, N, V)$, which may be obtained by comparison with the fundamental thermodynamic identity in the following way:

$$\Delta S = (\partial S/\partial U)_{N,V}\Delta U + (\partial S/\partial N)_{U,V}\Delta N + (\partial S/\partial V)_{U,N}\Delta V = (1/T)\Delta U - (\mu_c/T)\Delta N + (p/T)\Delta V$$

$$\rightarrow \quad \Delta U = (\partial U/\partial S)_{N,V}\Delta S + (\partial U/\partial N)_{S,V}\Delta N + (\partial U/\partial V)_{S,N}\Delta V = T\Delta S + \mu_c \Delta N - p\Delta V. \tag{57}$$

The above equation should be interpreted as follows. When the number of particles $N$ is constant, the change of the internal energy $\Delta U$ is equal to the amount of heat $\Delta Q = T\Delta S$ added to the system minus the amount of work $\Delta W = p\Delta V$ delivered to the outside world. However, if the number of particles changes by $\Delta N$, the change $\Delta S$ of the entropy is no longer equal to $\Delta Q/T$. In this case, we must correct $T\Delta S$ by the addition of $\mu_c \Delta N$ in order to account for the change of the entropy in consequence of the addition of new particles to the system.

From Eq.(57) it is seen that $(\partial U/\partial S)_{N,V} = T$, in agreement with the first identity in Eq.(48). Also, $(\partial U/\partial N)_{S,V} = \mu_c$ and $(\partial U/\partial V)_{S,N} = -p$, but these latter thermodynamic relations cannot be derived from the fundamental thermodynamic identity by a simple division and multiplication by $\partial S$ on the left-hand-side of each equation. In other words, attention must be paid to what is being varied and what is being kept constant. For example, Eq.(49) shows that, when $S$ and $N$ are fixed, it is necessary to have $U^{3/2}V$ = constant. Differentiation with respect to $V$ then yields

$$\tfrac{3}{2}(\partial U/\partial V)_{S,N}U^{1/2}V + U^{3/2} = 0 \quad \rightarrow \quad (\partial U/\partial V)_{S,N} = -\tfrac{2}{3}(U/V) = -Nk_B T/V = -p. \tag{58}$$

To show that $(\partial U/\partial N)_{S,V} = \mu_c$, one may use Eq.(49) to solve the following equation for $\Delta U$:

$$(\Delta S)_V = S(U + \Delta U, N + 1, V) - S(U, N, V) = 0. \tag{59}$$

Considering that, in the above equation, $\Delta N = 1$, it is clear that the solution must be $\Delta U = \mu_c$. An alternative, although ultimately equivalent, method of arriving at the same result is via partial differentiation, as follows:

$$(\Delta S)_V \cong (\partial S/\partial U)_{N,V}\Delta U + (\partial S/\partial N)_{U,V}\Delta N = (1/T)\Delta U - (\mu_c/T)\Delta N = 0$$

$$\rightarrow \quad (\Delta U/\Delta N)_{S,V} \cong \mu_c \quad \rightarrow \quad \mu_c = (\partial U/\partial N)_{S,V}. \tag{60}$$

**Example 4**. Consider an $N$-particle ideal gas initially held at temperature $T$ and volume $V$. In a reversible, adiabatic expansion, no heat is given to or taken from the gas, while the work delivered to the outside world by the pressure of the expanding gas causes the internal energy $U$ and, consequently, the temperature $T$ to drop. Since the change of entropy $\Delta S$ in a reversible process at temperature $T$ is equal to $\Delta Q/T$, and since, in the present adiabatic process, $\Delta Q$ is zero at all temperatures, we conclude that the entropy $S$ has remained constant. We write

$$\Delta W = p\Delta V = \tfrac{2}{3}(U/V)\Delta V = -\Delta U \quad \rightarrow \quad \frac{dU}{U} = -\frac{2}{3}\frac{dV}{V} \quad \rightarrow \quad \int \frac{dU}{U} = -\frac{2}{3}\int \frac{dV}{V}$$

$$\rightarrow \quad \ln(U_2/U_1) = -\tfrac{2}{3}\ln(V_2/V_1) \quad \rightarrow \quad U_2/U_1 = (V_1/V_2)^{2/3}$$

$$\rightarrow \quad T_2/T_1 = (V_1/V_2)^{2/3} \quad \rightarrow \quad TV^{2/3} = \text{constant}. \tag{61}$$

This is consistent with the Sackur-Tetrode equation, Eq.(44), which shows that the entropy does not change when both $N$ and $TV^{2/3}$ are kept constant. Another consequence of Eq.(61) is obtained by multiplying both sides of the equation with $Nk_B$, then using the ideal gas law to replace $Nk_BT$ with $pV$. We will have

$$pV^{5/3} = \text{constant}. \tag{62}$$

This is a well-known relation between the pressure and volume of an ideal gas under adiabatic expansion (or compression). More generally, this relation is written as $pV^\gamma = $ constant, where, for imperfect gases, the exponent $\gamma$ is below $5/3$. The corresponding relation between pressure and energy-density for a non-ideal gas is $pV = (\gamma - 1)U$, indicating that only a fraction of the total energy of the atoms/molecules belongs to the kinetic energy of translational motion, which is directly associated with the pressure of the gas.

**Concluding remarks**. Invoking several thermodynamic identities, we have derived and analyzed certain properties of blackbody radiation that were previously discussed, albeit from a different perspective, in an earlier paper [1]. We examined the sources of various contributions to energy fluctuations, the connection between blackbody radiation and the Johnson noise in electronic circuits, and the intimate relation between the statistics of photo-electron counts in photodetection and the random variations of the energy-density of thermal radiation. In the latter part of the paper we derived expressions for the entropy, energy-density, pressure, and chemical potential of an ideal monatomic gas under the conditions of thermal equilibrium. Our goal here has been to compare and contrast the thermodynamic properties of the photon gas (i.e., blackbody radiation) against those of a rarefied gas consisting of rigid, identical particles of matter.